\documentclass[intlimits,twoside,a4paper]{article}

\usepackage{amsmath,amssymb}
\usepackage{graphicx}
\usepackage[T2A]{fontenc}
\usepackage[cp1251]{inputenc}

\usepackage{cmpj2}
\issue{2013}{16}{2}{23003}

\doinumber{10.5488/CMP.16.23003}

\title[Wilson's ERGE and anomalous dimension]%
{The Wilson exact renormalization group equation \\
and the anomalous dimension parameter }%
\author[C. Bervillier]{C. Bervillier\thanks{E-mail: claude.bervillier@lmpt.univ-tours.fr}}
\address{
Laboratoire de Math\'{e}matiques et Physique Th\'{e}orique,
UMR 7350 (CNRS), F\'{e}d\'{e}ration Denis Poisson,\\
Universit\'{e} Fran\c{c}ois Rabelais, Parc de Grandmont, 37200 Tours, France}

\date{Received December 14, 2012, in final form March 8, 2013}
\authorcopyright{C. Bervillier, 2013}

\begin{document}

\maketitle

\begin{abstract}
The non-linear way the anomalous dimension parameter has been introduced in
the historic first version of the exact renormalization group equation is
compared to current practice. A simple expression for the exactly marginal
redundant operator proceeds from this non-linearity, whereas in the linear
case, first order differential equations must be solved to get it. The role
of this operator in the construction of the flow equation is highlighted.
\pacs 05.10.Cc, 11.10.Gh, 64.60.ae
\end{abstract}

\section{Introduction}

The Wilson renormalization group (RG) ideas \cite{440,4011,2835} have
formalized and clarified the notions of scaling and of universality attached
to critical behavior. As several modern subjects in physics, critical
phenomena cannot be studied by pure perturbative methods. The development of
nonperturbative methods is thus very important. As is well demonstrated by
the work of M.~Kozlovskii (see, e.g., \cite{6707,5175,6071,6715,7848}), the
collective variables method \cite{7842}, close to the early ideas of
Kadanoff \cite{248}, is one of the RG-like nonperturbative methods allowing
one to investigate the critical point, starting with a microscopic Hamiltonian~--- thus allowing one to calculate nonuniversal quantities such as the
critical temperature \cite{6707}. The subject of the present article is
another expression of the nonperturbative RG framework, called the exact RG
equation (ERGE)~--- see in \cite{440} section 11, equations~(11.14, 11.15), and
for reviews see, e.g., \cite{4823,4595,4700,5141,5604,6022,6156,6723,6699}.
Though one could also calculate nonuniversal quantities with an ERGE (see,
e.g., \cite{4245,4716,6709}), our purpose is presently limited to the close
vicinity of a RG fixed point with a view to discuss the particular issue
of the way one may account for the anomalous dimension of the field
(the critical exponent $\eta $) within an ERGE.

Among an infinite variety of equivalent ERGE, only those based on an
effective cutoff function associated with a bilinear kinetic term are
currently used \cite{4753}. Despite this limited number of usual variants (a
priori close to each other), they show  differences sufficient to make it
sometimes difficult to clearly display (or even understand) their
relationships. Of course, several authors have addressed this issue in
general \cite{3358,2520,3550,4619,3357,3993,4436} but, in contrast to \cite%
{5744,6750,7289,7205}, the anomalous dimension parameter $\eta $ was not
included in their considerations. For example, it is known that the Wilson
historic first version \cite{440} is made equivalent to the Polchinski
version \cite{354}, provided there is a choice of an exponential cutoff function and
after a specific field redefinition \cite{3357,5744}, but the parameter $%
\eta $ was not explicitly considered in this relation except in \cite{5744}.

Recently Osborn and Twigg \cite{7289} and Rosten \cite{6750,7205} have
independently established the relation between a version of the Polchinski
ERGE~--- ``modified'' to include the
anomalous dimension \cite{3491}~--- and a ``fixed-point
equation'' for the effective average action \cite{3357}.
This relation appears to be rather complicated:
it ``involves solving some first order differential equations''
\cite{7289}. O'Dwyer and Osborn \cite{6228} had previously encountered
similar first order differential equations in the process of constructing
the (exactly) marginal, redundant ``operator''\footnote{
Actually $\mathcal{O}$ is not an operator but represents a direction in the
space of actions $\left\{ S\right\} $.} $\mathcal{O}$ (EMRO) associated with
the change of normalization of the field by a constant factor but they did
not discuss the effect of the way $\eta $ had been accounted for in the
ERGE on this complexity. Yet, it is a well known fact that the EMRO\ of the
Wilson version \cite{440} takes on a simple form~--- e.g., see equation~(24) of
\cite{4595} or equation~(2.35) of \cite{4405}. The object of the present article
is to compare the main ways of introducing $\eta $ in an ERGE encountered in
the literature. We show, in particular, that the simplicity of the form of
the EMRO is maintained in the Wilson version of the ERGE extended to an
arbitrary cutoff function \cite{5744}. The role of the EMRO in the
construction of an ERGE is also highlighted. A larger discussion is left to
an upcoming detailed publication \cite{Berv}.

\section{The RG steps}

\subsection{Reminder\label{reminder}}

Basically, a RG transformation of a cutoffed action $S[\phi ,\ell ]$ of a
scalar field $\phi $ involves three steps. Having chosen a fixed arbitrary
momentum scale of reference $\Lambda _{0}$, and having defined the RG-scale
parameter $\ell ={\Lambda }/{\Lambda _{0}}\leqslant 1$, these three steps are:\\[-7mm]

\parbox[t]{14.4cm}{

\begin{enumerate}
\item[Step 1:] an integration of the high momentum components of the field
generating an effective action with a reduced cutoff $\Lambda ^{\prime
}=\left( 1-{d\ell }/{\ell }\right) \Lambda $.

\item[Step 2:] a rescaling of the momenta back to the initial value of the
cutoff: $\left\vert q\right\vert \rightarrow \left( 1-{d\ell }/{\ell }%
\right) \left\vert q\right\vert $. This step is accounted for by a simple
dimensional analysis such as defining a dimensionless momentum as $\mathbf{%
\tilde{q}=q}/\Lambda $ and a dimensionless field $\tilde{\phi}_{\tilde{q}%
}=\Lambda ^{-\bar{d}_{\phi }}\phi _{q}\,$\ [see also equation~(\ref%
{eq:dphibardphi})].

\item[Step 3:] a renormalization of the field $\tilde{\phi}_{\tilde{q}%
}\longrightarrow \zeta \left( \ell -d\ell \right) \,\tilde{\phi}_{\tilde{q}}$%
, with $\zeta \left( \ell \right) =\ell ^{\varpi }$ and $\varpi $ related to
$\eta $, e.g., see equation~(\ref{eq:omega}). This step is required to keep
constant (i.e., independent of $\ell $) one term of $S$ in order to set up
the system of scale of reference.
\end{enumerate}
}

Having defined the RG-time $t$ as:%
\begin{equation}
t=-\ln \ell =-\ln \left( \frac{\Lambda }{\Lambda _{0}}\right) \,,
\label{eq:RG-time}
\end{equation}%
the combined effect of the above three steps on the action $S$ may be
expressed as:
\begin{equation}
\dot{S}[\tilde{\phi},e^{-t}]=\mathcal{G}_{\mathrm{tra}}\left( S\right) +%
\mathcal{G}_{\mathrm{dil}}\left( S,d_{\phi }\right) +\mathcal{G}_{\mathrm{ren%
}}\left( S\right) \,,  \label{eq:SpointGene}
\end{equation}%
in which:%
\begin{eqnarray}
\dot{S}[\tilde{\phi},\re^{-t}] &\equiv &\left. \frac{\rd}{\rd t}S[\tilde{\phi}%
,\re^{-t}]\right\vert _{\tilde{\phi}}\,,  \label{eq:Spoint} \\
\mathcal{G}_{\mathrm{dil}}\left( S,d_{\phi }\right)  &=&\int_{\tilde{p}%
}\left( \mathbf{\tilde{p}}\cdot \frac{\partial \tilde{\phi}_{\tilde{p}}}{%
\partial \mathbf{\tilde{p}}}-\bar{d}_{\phi }\tilde{\phi}_{\tilde{p}}\right)
\frac{\delta S}{\delta \tilde{\phi}_{\tilde{p}}}\,.  \label{eq:Gdil}
\end{eqnarray}

The expression of $\mathcal{G}_{\mathrm{tra}}\left( S\right) $ in (\ref%
{eq:SpointGene}), depends on the way the above step 1 is realized. In any
case, for a complete action $S$, it corresponds to a (quasi local) field
redefinition which leaves the partition function unchanged and,
consequently, must have the form \cite{4011,2835}:%
\begin{eqnarray}
\mathcal{G}_{\mathrm{tra}}\left( S\right)  &=&\Xi \left[ S,\Psi \right] \,,
\label{eq:gtra2} \\
\Xi \left[ S,\Psi \right]  &=&\int_{\tilde{q}}\left( \Psi _{\tilde{q}}\frac{%
\delta S}{\delta \tilde{\phi}_{\tilde{q}}}-\frac{\delta \Psi _{\tilde{q}}}{%
\delta \tilde{\phi}_{\tilde{q}}}\right) \,,  \label{eq:GtraGene}
\end{eqnarray}%
in which $\Psi _{\tilde{q}}$ is some functional of $\tilde{\phi}$.

As for the expression of $\mathcal{G}_{\mathrm{ren}}\left( S\right) $, it is
the object of the present article to discuss its various forms encountered
in the literature. It is related to the expression of $\bar{d}_{\phi }$ in (%
\ref{eq:Gdil}); if $d_{\phi }$ is the dimension of $\ \phi \left( x\right) $%
, $\bar{d}_{\phi }$ is that of its Fourier transformed $\phi _{q}$, if $D$
is the spatial dimension, they are related through:%
\begin{equation}
\bar{d}_{\phi }=d_{\phi }-D\,.  \label{eq:dphibardphi}
\end{equation}%
In the following we use two kinds of $d_{\phi }$: a ``classical'' family $d_{\phi }^{\left( c\right) }$:%
\begin{equation}
d_{\phi }^{\left( c\right) }=\frac{D}{2}-n_{0}\,,  \label{eq:dphiWil}
\end{equation}%
in which\footnote{%
The value of $d_{\phi }^{(\mathrm{c})}$ depends on which term is chosen to set the
standard of the scale of reference. Usually one refers to the kinetic term
and $d_{\phi }^{\left(\mathrm{c}\right) }=\frac{D-2}{2}$ ($n_{0}=1$). Wilson has
chosen a pure quadratic term and, for him $d_{\phi }^{\left(
\mathrm{c}\right) }=\frac{D}{2}$ ($n_{0}=0$).} $n_{0}$ is related to the behavior of
the cutoff function at small momenta [see equations~(\ref{eq:FormeDeP}, \ref%
{eq:FormeDePK}) where $n_{0}=0$ or $n_{0}=1$] , and the fixed-point-value
dimension:%
\begin{equation}
d_{\phi }^{\left( a\right) }=\frac{D-2+\eta }{2}\,,  \label{eq:dphia}
\end{equation}%
in which $\eta $ is the usual critical exponent.

\subsection{Main ways of introducing $\protect\eta $}

Though it gave the name to the theory, step 3 is currently skipped\ in the
modern constructions of an ERGE. A perfect example of this fact is the
original Polchinski ERGE \cite{354} a version of which~--- including $\eta $
as proposed in \cite{3491}~--- has been called the ``modified'' Polchinski version \cite{6750,7205} or the\
Polchinski version ``extended to include the parameter $\eta
$'' \cite{7289}. In modern approaches, the renormalization
step 3 is currently avoided and $\eta $ is merely included in the rescaling
step 2 via some ad hoc ``anomalous''
dimensional analysis \cite{3357,3912}. There is nothing wrong with such a
procedure: the large freedom that is offered in the construction of an ERGE
makes it well acceptable. It remains no less true that it is fair to ask
whether or not there are important consequences of doing this or that.

Essentially, two ways of introducing $\eta $ are encountered in the
literature:

\begin{enumerate}
\item via the explicit renormalisation step 3 associated with $\mathcal{G}_{%
\mathrm{ren}}\left( S\right) $. In that case the dimension\ $d_{\phi }$ in $%
\mathcal{G}_{\mathrm{dil}}\left( S,\ d_{\phi }\right) $ is classical, given
by (\ref{eq:dphiWil}). This is the original and most general procedure but it is
not currently used.

\item via some \textquotedblleft anomalous\textquotedblright\ dimensional
analysis as proposed, e.g., by Morris \cite{3357}. In that case $\mathcal{G}%
_{\mathrm{ren}}\left( S\right) \equiv 0$ and\ $d_{\phi }$ is given by (\ref%
{eq:dphia}). In fact, this procedure, or some equivalent variant, is
generally used in the current versions of the ERGE (see also, e.g., Berges
et al \cite{4700}, Ball et al \cite{3491} etc. \ldots). It is a correct procedure
in case of the proximity of a fixed point.
\end{enumerate}

In addition, one may observe that the effective contribution of $\eta $
within the ERGE is either purely linear or not (w.r.t. $S$).\ For example,
the procedure of Ball et al \cite{3491} is linear whereas the Wilson \cite%
{440}, Morris \cite{3357} and Wetterich \cite{4700} procedures are
non-linear. These differences, although well allowed, have some consequences
in the relations between the main versions of the ERGE. In the following
sections we examine the non-linear Wilson procedure in greater detail than
previously done in \cite{5744} and compare it with the linear and non-linear
procedures of, respectively, Ball et al \cite{3491} and Morris \cite{3357}.

\section{The extended Wilson ERGE\label{sec:EWE}}

\subsection{Presentation}

Let us consider the extended Wilson ERGE (i.e., the flow equation of $S$ with
an arbitrary cutoff function), as obtained in \cite{5744}. For convenience,
we adopt the notations of Osborn and Twigg \cite{7289} and write the flow
equation under the form of (\ref{eq:SpointGene}) with (notice the change in $%
\mathcal{G}_{\mathrm{dil}}$):%
\begin{eqnarray}
\mathcal{G}_{\mathrm{tra}}\left( S\right)  &=&\int_{\tilde{q}}\left[ G\left(
\tilde{q}^{2}\right) \left( \frac{\delta ^{2}S}{\delta \tilde{\phi}_{\tilde{q%
}}\delta \tilde{\phi}_{-\tilde{q}}}-\frac{\delta S}{\delta \tilde{\phi}_{%
\tilde{q}}}\frac{\delta S}{\delta \tilde{\phi}_{-\tilde{q}}}\right) +H\left(
\tilde{q}^{2}\right) \tilde{\phi}_{\tilde{q}}\frac{\delta S}{\delta \tilde{%
\phi}_{\tilde{q}}}\right] \,,  \label{eq:Gtra} \\
\mathcal{G}_{\mathrm{ren}}\left( S\right)  &=&\varpi \,\mathcal{O}( S,%
\tilde{P}) \,,  \label{eq:Gren2} \\
\mathcal{G}_{\mathrm{dil}}\left( S,d_{\phi }\right)  &=&\mathcal{G}_{\mathrm{%
dil}}( S,d_{\phi }^{\left( c\right) }) \,,  \label{eq:Gdil2}
\end{eqnarray}%
where:%
\begin{eqnarray}
\mathcal{O}( S,\tilde{P})  &=&\int_{\tilde{q}}\left[ \tilde{P}%
\left( \tilde{q}^{2}\right) \left( \frac{\delta ^{2}S}{\delta \tilde{\phi}_{%
\tilde{q}}\delta \tilde{\phi}_{-\tilde{q}}}-\frac{\delta S}{\delta \tilde{%
\phi}_{\tilde{q}}}\frac{\delta S}{\delta \tilde{\phi}_{-\tilde{q}}}\right) +%
\tilde{\phi}_{\tilde{q}}\frac{\delta S}{\delta \tilde{\phi}_{\tilde{q}}}%
\right] \,,  \label{eq:O1} \\
G\left( \tilde{q}^{2}\right)  &=&-\tilde{q}^{2}\tilde{P}\left( \tilde{q}%
^{2}\right) \frac{K^{\prime }\left( \tilde{q}^{2}\right) \,}{K\left( \tilde{q%
}^{2}\right) },  \label{eq:Gq2} \\
H\left( \tilde{q}^{2}\right)  &=&-2\tilde{q}^{2}\frac{K^{\prime }\left( \tilde{q}^{2}\right) \,}{K\left( \tilde{q%
}^{2}\right) } \,.  \label{eq:Hq2}
\end{eqnarray}

Compared to \cite{5744}, the cutoff function $P\left( q^{2},\Lambda \right) $
has been given the (regular) dimension $\dim \left[ P\right] =-2n_{0}$ so
that:%
\begin{eqnarray}
P(q^{2},\ell \Lambda _{0}) &=&\Lambda _{0}^{-2n_{0}}\ell ^{2\varpi -2n_{0}}%
\tilde{P}\left( \tilde{q}^{2}\right) \,,  \label{eq:FormeDeP} \\
\tilde{P}\left( \tilde{q}^{2}\right)  &=&\frac{K\left( \tilde{q}^{2}\right)
}{\left( \tilde{q}^{2}\right) ^{n_{0}}}\,,  \label{eq:FormeDePK}
\end{eqnarray}%
where $K$ is dimensionless~--- this is to permit comparison with the current uses
where $n_{0}=1$ whereas $n_{0}=0$ in \cite{5744}.

It is useful to recall the following points:

\begin{itemize}
\item the general form of an ERGE involves an additive field independent
term which will be sytematically neglected in the following;

\item the establishment of the extended Wilson ERGE is based on the relation
of the complete action $S$ to a partial action $S_{\mathrm{int}}$ by
extracting a quadratic form involving the arbitrary cutoff function $P$:
\begin{equation}
S\left[ \phi \right] =\frac{1}{2}\int_{q}\phi _{q}P^{-1}(q^{2},\ell \Lambda
_{0})\phi _{-q}+S_{\mathrm{int}}\left[ \phi \right] \,;
\label{eq:Polch-Action}
\end{equation}

\item in its original version \cite{354} Polchinski expresses the flow of $%
S_{\mathrm{int}}$ under a change of $\ell $~--- instead of $S$ for the Wilson
ERGE;

\item the factorized $\ell ^{2\varpi }$-term in (\ref{eq:FormeDeP}) was not
part of Polchinski's assumptions. This kind of $\ell $-dependency in front
of the cutoff function is a convenient artefact to introduce $\eta $
non-linearly in the ERGE. This is not unusual since Morris \cite{3357} already
used it when he gave the cutoff function an anomalous dimension (see section %
\ref{sec:MorrisVsWilson}). We shall show that the recourse to the EMRO
enables us to get rid of this artefact;

\item in order to specify the nature of the renormalization step, one must
make reference to the right power law behavior at large distances of the
critical two point correlation function and this implies that \cite{5744}:%
\begin{equation}
\varpi =1-n_{0}-\frac{\eta }{2}\,.  \label{eq:omega}
\end{equation}%
[In the case of anomalous dimensional analysis, this step is not
required since the field has a priori been given  the \textquotedblleft
right\textquotedblright\ fixed point dimension $d_{\phi }^{\left( a\right) }$
defined by (\ref{eq:dphia})];

\item it is easy to verify that for $n_{0}=0$ and $\varpi $ given by (\ref%
{eq:omega}), (\ref{eq:Gtra})--(\ref{eq:FormeDePK}) gives the historic
first version of the ERGE \cite{440} with the choice $\tilde{P}(\tilde{q}%
^{2})=\re^{-2\tilde{q}^{2}}$, and the redefinition of the field \cite{5744}:
\begin{equation}
\tilde{\phi}_{\tilde{q}}\rightarrow \sqrt{\tilde{P}(\tilde{q}^{2})}\,\tilde{%
\phi}_{\tilde{q}}\,;  \label{eq:Change}
\end{equation}

\item the writing of equations~(\ref{eq:Gtra})--(\ref{eq:Hq2}) suggests that $%
\mathcal{O}( S,\tilde{P}) $ plays the role of the redundant
\textquotedblleft operator\textquotedblright\ $\mathcal{O}\left( S,1\right) $
which is associated with an infinitesimal renormalization of the field by a
constant factor \cite{4421,4405}. It is, actually, via\ $\mathcal{O}\left(
S,1\right) $ that $\eta $ was introduced in the historic first version%
\footnote{%
Via a scale dependent factor $\varpi \left( t\right) $ in front of \ $%
\mathcal{O}\left( S,1\right) $ instead of a constant $\varpi $ as in (\ref%
{eq:Gren2}), $\varpi \left( t\right) $ coincides with $\varpi $ in the
vicinity of a fixed point.}.
\end{itemize}

Instead of assuming the presence of the factor $\ell ^{2\varpi }$ in (\ref%
{eq:FormeDeP}), we could as well have implemented step 3 using $\mathcal{O}%
\left( S,1\right) $ in place of $\mathcal{O}( S,\tilde{P}) $ in (%
\ref{eq:Gren2}), but then this would have destroyed the currently admitted
equivalence \cite{3357,4753} [under the change (\ref{eq:Change})] between
Polchinski's and Wilson's ERGE. This is merely because, given a fixed point $%
S^{\ast }$ [characterized by $\dot{S}^{\ast }=0$], $\mathcal{O}(
S^{\ast },1) $ is not an EMRO for the extended Wilson ERGE whereas $%
\mathcal{O}( S^{\ast },\tilde{P}) $ is. It is the object of the
next section to discuss this issue.

\subsection{The exactly marginal redundant \textquotedblleft
operator\textquotedblright}

In this section, we apply  the procedure described
in appendix D of \cite{6228} to the complete action $S$ (see also \cite{6699}) and we demonstrate that $%
\mathcal{O}( S,\tilde{P}) $ defined by (\ref{eq:O1}) with $%
S=S^{\ast }$, corresponds to an EMRO for the ERGE given by (\ref%
{eq:SpointGene}), (\ref{eq:Gtra})--(\ref{eq:Gdil2}), and that it may be used to
implement the RG step~3.

It is easy to see that $\mathcal{O}( S,\tilde{P}) $ is a
redundant \textquotedblleft operator\textquotedblright\ since it may be
written under the form of the r.h.s. of (\ref{eq:GtraGene}) with:%
\begin{equation}
\Psi _{\tilde{q}}^{\left( \tilde{P}\right) }=\tilde{\phi}_{\tilde{q}}-\tilde{%
P}\left( \tilde{q}^{2}\right) \frac{\delta S}{\delta \tilde{\phi}_{-\tilde{q}%
}}\,.  \label{eq:PsiqP}
\end{equation}

\parbox[t]{14.4cm}{
We then proceed in two stages, showing successively that:\\[-5mm]

\begin{enumerate}
\item [\hfill{Stage~1:}]  whatever the functions $G\left( \tilde{q}^{2}\right) $ and $%
H\left( \tilde{q}^{2}\right) $:
\begin{enumerate}
\item[(a)] there exists a function $B\left( q^{2}\right) $~--- solution of a first
order differential equation~--- such that $\mathcal{O}\left( S^{\ast
},B\right) $ is exactly marginal for the ERGE without renormalization (i.e.,
with $\mathcal{G}_{\mathrm{ren}}\left( S\right) \equiv 0$ and an arbitrary
dimension parameter $d_{\phi }$),
\item[(b)] $\mathcal{O}\left( S^{\ast },B\right) $ is again exactly marginal for
the ERGE translated by $\alpha \,\mathcal{O}\left( S,B\right) $ [i.e., with $%
\mathcal{G}_{\mathrm{ren}}\left( S\right) =\alpha \,\mathcal{O}\left(
S,B\right) $] with $\alpha $ an arbitrary constant.
\end{enumerate}
\item[{Stage~2:}]
$B\left( q^{2}\right) $ identifies with $\tilde{P}\left(
\tilde{q}^{2}\right) $ in the case of (\ref{eq:Gq2}), (\ref{eq:Hq2}).
\end{enumerate}
}

Let us consider the ERGE linearized about a fixed point $S^{\ast }$ and its
associated eigenvalue equation for some eigenfunctional $\Theta _{i}[
\tilde{\phi}] $:
\begin{equation}
\mathcal{\hat{D}}\,\Theta _{i}=\lambda _{i}\Theta _{i}\,.  \label{eq:vp}
\end{equation}

For the ERGE given by (\ref{eq:SpointGene}), (\ref{eq:Gtra})--(\ref{eq:Gdil2}),
without the renormalisation part $\mathcal{G}_{\mathrm{ren}}\left( S\right) $
and for arbitrary $d_{\phi }$, we have:%
\begin{eqnarray*}
\mathcal{\hat{D}} &=&\mathcal{\hat{D}}_{1}+\mathcal{\hat{D}}_{2}+\mathcal{%
\hat{D}}_{3}+\mathcal{\hat{D}}_{4}\,, \\
\mathcal{\hat{D}}_{1} &=&\int_{\tilde{p}}\left[ \mathbf{\tilde{p}}\cdot
\frac{\partial \tilde{\phi}_{\tilde{p}}}{\partial \mathbf{\tilde{p}}}\,-\bar{%
d}_{\phi }\,\tilde{\phi}_{\tilde{p}}\right] \frac{\delta }{\delta \tilde{\phi%
}_{\tilde{p}}}\,, \\
\mathcal{\hat{D}}_{2} &=&\int_{\tilde{p}}G\left( \tilde{p}^{2}\right) \frac{%
\delta ^{2}}{\delta \tilde{\phi}_{\tilde{p}}\delta \tilde{\phi}_{-\tilde{p}}}%
\,, \\
\mathcal{\hat{D}}_{3} &=&\int_{\tilde{p}}H\left( \tilde{p}^{2}\right) \,%
\tilde{\phi}_{\tilde{p}}\frac{\delta }{\delta \tilde{\phi}_{\tilde{p}}}\,, \\
\mathcal{\hat{D}}_{4} &=&-2\int_{\tilde{p}}G\left( \tilde{p}^{2}\right)
\frac{\delta S^{\ast }}{\delta \tilde{\phi}_{\tilde{p}}}\frac{\delta }{%
\delta \tilde{\phi}_{-\tilde{p}}}\,.
\end{eqnarray*}

Considering a general redundant \textquotedblleft
operator\textquotedblright\ $\Xi \left[ S,\Psi \right] $ as defined by (\ref%
{eq:GtraGene}), it is not very complicated to verify that the general
property \cite{4011,2835} of the redundant \textquotedblleft
operators\textquotedblright\ to form a closed subspace under the flow in the
vicinity of a fixed-point takes on the form:%
\begin{equation}
\mathcal{\hat{D}}\Xi [ S^{\ast },\Psi ] =\Xi [ S^{\ast },%
\mathcal{\hat{D}}_{t}\Psi ] \,,  \label{eq:Redondance}
\end{equation}%
with:%
\begin{equation}
\mathcal{\hat{D}}_{t}\Psi _{\tilde{q}}[ \tilde{\phi}] =\left(
\mathcal{\hat{D}}+\bar{d}_{\phi }-G\left( \tilde{q}^{2}\right) -\mathbf{%
\tilde{q}}\cdot \frac{\partial }{\partial \mathbf{\tilde{q}}}\right) \Psi _{%
\tilde{q}}[ \tilde{\phi}] \,.  \label{eq:Drondt}
\end{equation}

Following the procedure of \cite{6228}, we try to build up the EMRO [i.e., a
solution $\Theta _{0}$ of (\ref{eq:vp}) with $\lambda _{0}=0$] from equations~(\ref%
{eq:Redondance}), (\ref{eq:Drondt}) with a function $\Psi _{\tilde{q}}^{(B)}%
[ \tilde{\phi}] $ similar to (\ref{eq:PsiqP}) and find that $%
B\left( x\right) $ must be the solution of the following first order
differential equation (with initial value):%
\begin{eqnarray}
&&G\left( x\right) -\varpi _{g}B\left( x\right) -H\left( x\right) B\left(
x\right) -xB^{\prime }\left( x\right)  =0\,,  \label{eq:B} \\
&&B\left( 0\right)  =\frac{G\left( 0\right) }{\varpi _{g}}\,,
\label{eq:Binit}
\end{eqnarray}%
in which:%
\begin{equation}
\varpi _{g}=-\frac{d_{\phi }+\bar{d}_{\phi }}{2}\,.  \label{eq:omegaBarg}
\end{equation}

Assuming an explicit solution of $B\left( x\right) $ [see equations~(\ref{eq:BSol}), (\ref{eq:C0Sol})] this closes stage 1a.

The quantity $\mathcal{O}\left( S,B\right) ,$ similar to (\ref{eq:O1}), is
thus redundant and exactly marginal at the fixed point $S^{\ast }$. Let us
consider a new ERGE with:%
\begin{equation}
\mathcal{G}_{\mathrm{ren}}\left( S\right) =\alpha \,\mathcal{O}\left(
S,B\right) \,.  \label{eq:GrenB}
\end{equation}

We easily see that this modification amounts to merely having performed the
following translations in the flow equation:%
\begin{eqnarray}
G\left( x\right)  &\longrightarrow &G\left( x\right) +\alpha B \,,
\label{eq:translatAlpha1} \\
\varpi _{g} &\longrightarrow &\varpi _{g} + \alpha \,,
\label{eq:translatAlpha2}
\end{eqnarray}%
which keep equation (\ref{eq:B}) unchanged. Consequently, stage~1b is also
verified. It is clear that this would not be the case using $\mathcal{O}%
\left( S,1\right) $ in (\ref{eq:GrenB}).

The solution of the differential equations (\ref{eq:B}), (\ref{eq:Binit}) is:%
\begin{eqnarray}
B\left( x\right)  &=&\left( \frac{1}{x}\right) ^{\varpi _{g}}\frac{1}{%
C_{0}\left( x\right) }\int_{0}^{x}u^{\varpi _{g}-1}G(u)C_{0}\left( u\right)
\rd u\,,\qquad \text{if}\qquad \varpi _{g}\geqslant0\,\,,  \label{eq:BSol} \\
C_{0}\left( x\right)  &=&\exp\left\{\int_{0}^{x}\frac{H(u)}{u}\,\rd u\right\},\qquad\hspace{2.4cm} \text{if}%
\qquad H(0)=0\,.  \label{eq:C0Sol}
\end{eqnarray}%
Using an integration by parts, one may verify that $B\left( x\right) $, as
given by (\ref{eq:BSol}), satisfies the initial condition (\ref{eq:Binit}),
provided that $\varpi _{g}\geqslant0$.

In the case where the functions $G$ and $H$ are defined by (\ref{eq:Gq2}), %
(\ref{eq:Hq2}), equation~(\ref{eq:BSol}) simplifies to give:%
\begin{equation}
B\left( x\right) \equiv \tilde{P}\left( x\right) \,,  \label{eq:B=P}
\end{equation}%
so that with $\varpi _{g}=\varpi $, as given by (\ref{eq:omega}), the
quantity $\varpi \,\mathcal{O}( S,\tilde{P}) $ corresponds to the
effective non-linear realization of the field renormalization adapted to the
ERGE under consideration (QED).

In order to discuss the linear implementation of $\eta $, it is useful to
first consider the relation back to the original Polchinski version of the
ERGE.

\section{From Wilson to Polchinski}

In this section we scrupulously  look at the relation between the Wilson and the
Polchinski ERGE considering successively the linear (Ball et al \cite{3491})
and non-linear (Wilson \cite{440,5744}, Morris \cite{3357}) ways of
introducing $\eta $.

The Polchinski equation for $S_{\mathrm{int}}$ \cite{354} is originally
limited to the implementation of the RG step~1 via the variation of $S_{%
\mathrm{int}}$ under the change of $\Lambda $ introduced by a bilinear term
in $S$ [as shown in (\ref{eq:Polch-Action})]. It reads:%
\begin{equation}
\left. \ell \frac{\partial }{\partial \ell }S_{\mathrm{int}}\right\vert
_{\phi }=-\frac{1}{2}\int_{q}\left. \ell \frac{\partial }{\partial \ell }%
P\left( q^{2},\ell \Lambda _{0}\right) \right\vert _{q}\left[ \frac{\delta
^{2}S_{\mathrm{int}}}{\delta \phi _{q}\delta \phi _{-q}}-\frac{\delta S_{%
\mathrm{int}}}{\delta \phi _{q}}\frac{\delta S_{\mathrm{int}}}{\delta \phi
_{-q}}\right] \,.  \label{eq:PolInt}
\end{equation}

From this incomplete expression of the RG flow, the shift to $S$ is easily
implemented via the following functional derivatives of (\ref%
{eq:Polch-Action}):%
\begin{eqnarray}
\frac{\delta S_{\mathrm{int}}}{\delta \phi _{q}} &=&\frac{\delta S}{\delta
\phi _{q}}-P^{-1}\phi _{-q}\,\,,  \label{eq:dSdPhi} \\
\frac{\delta ^{2}S_{\mathrm{int}}}{\delta \phi _{q}\delta \phi _{-q}} &=&%
\frac{\delta ^{2}S}{\delta \phi _{q}\delta \phi _{-q}}-P^{-1}\,,
\label{eq:d2Sd2Phi}
\end{eqnarray}%
from which we get (up to an additive field-independent term):%
\begin{equation}
\left. \ell \frac{\partial }{\partial \ell }S\left[ \phi \right] \right\vert
_{\phi }=-\frac{1}{2}\left\{ \int_{q}\left. \ell \frac{\partial }{\partial
\ell }P\left( q^{2},\ell \Lambda _{0}\right) \right\vert _{q}\left[ \frac{%
\delta ^{2}S}{\delta \phi _{q}\delta \phi _{-q}}-\frac{\delta S}{\delta \phi
_{q}}\frac{\delta S}{\delta \phi _{-q}}+2P^{-1}\left( q^{2},\ell \Lambda
_{0}\right) \phi _{q}\frac{\delta S}{\delta \phi _{q}}\right] \right\} \,.
\label{eq:PolTot}
\end{equation}

Then, depending on the procedure chosen for introducing $\eta $, the
implementation of the RG steps 2 and 3 may be easier to get a hand on
according to whether one considers (\ref{eq:PolInt}) or (\ref{eq:PolTot}).
In the linear case of Ball et al \cite{3491} this does not matter, however.

\subsection{Linear introduction of $\protect\eta $}

The cumulative account from $\Lambda _{0}$ to $\Lambda =\ell \,\Lambda _{0}$
which has been conveniently included as an explicit $\ell $-factor within
the cutoff function [as displayed in (\ref{eq:FormeDeP})] is not necessary
to the construction of an ERGE . Actually, only the reduction of the degrees
of freedom in the infinitesimal range $\left[ \Lambda ,\left( 1-\frac{d\ell
}{\ell }\right) \Lambda \right] $ is obligatory. In other terms: the
renormalization step 3 may be implemented linearly only. This is the
procedure adopted by Ball et al \cite{3491} when they introduced $\eta $.
However, there is a supplementary freedom.

The role of the renormalization step 3 is to compensate the modification of
one term of $S$ induced by the implementation of the RG step 1. Instead of
explicitly renormalizing the field, one may as well view this modification
as an integral part of $\phi $ so that it may be absorbed in the rescaling
step 2 by assuming that $\phi $ has the (anomalous) dimension (\ref{eq:dphia}%
)~--- notice that this implies $n_{0}=1$. Ball et al \cite{3491} have chosen
that possibility. For both $S$ and $S_{\mathrm{int}}$, RG steps 2 and 3 are
then implemented by a mere dimensional analysis that readily provides the
final flow equations. In terms of $t$ defined in (\ref{eq:RG-time}) we get:

\begin{itemize}
\item the flow equation for $S$ under the form (\ref{eq:SpointGene}) with $%
\mathcal{G}_{\mathrm{ren}}\left( S\right) =0$, $\mathcal{G}_{\mathrm{dil}%
}\left( S,d_{\phi }\right) =\mathcal{G}_{\mathrm{dil}}( S,d_{\phi
}^{\left( a\right) }) $ and $\mathcal{G}_{\mathrm{tra}}\left( S\right)
$ given by (\ref{eq:Gtra}), (\ref{eq:Gq2}), (\ref{eq:Hq2});

\item the flow equation for $S_{\mathrm{int}}$ as:%
\begin{eqnarray}
\dot{S}_{\mathrm{int}} &=&\mathcal{G}_{\mathrm{tra}}\left( S_{\mathrm{int}%
}\right) +\mathcal{G}_{\mathrm{dil}}( S_{\mathrm{int}},d_{\phi
}^{\left( a\right) }) \,,  \label{eq:SptIntLin} \\
\mathcal{G}_{\mathrm{tra}}\left( S_{\mathrm{int}}\right)  &=&\int_{\tilde{q}}%
\left[ G\left( \tilde{q}^{2}\right) \left( \frac{\delta ^{2}S_{\mathrm{int}}%
}{\delta \tilde{\phi}_{\tilde{q}}\delta \tilde{\phi}_{-\tilde{q}}}-\frac{%
\delta S_{\mathrm{int}}}{\delta \tilde{\phi}_{\tilde{q}}}\frac{\delta S_{%
\mathrm{int}}}{\delta \tilde{\phi}_{-\tilde{q}}}\right) \right]
\label{eq:GtraIntLin}
\end{eqnarray}%
with $G\left( \tilde{q}^{2}\right) $ given by (\ref{eq:Gq2}).
\end{itemize}

However, since the anomalous dimension attributed to $\phi $ is not
compensated by an anomalous dimension of the cutoff function in the bilinear
term of $S$, the relation between the two flow equations is altered and we
have:%
\begin{equation}
\dot{S}=\dot{S}_{\mathrm{int}}+\varpi \int_{\tilde{q}}\tilde{\phi}_{\tilde{q}%
}\tilde{P}^{-1}\left( \tilde{q}^{2}\right) \tilde{\phi}_{-\tilde{q}}\,,
\label{eq:SptToSinptLinear}
\end{equation}%
with $\varpi $ given by (\ref{eq:omega}) in which $n_{0}=1$. This implies
that the two flow equations do not have equivalent fixed points.

Equation (\ref{eq:SptToSinptLinear}) has induced the notion of \textquotedblleft
modified\textquotedblright\ Polchinski flow equation \cite{7289,6750,7205}
which refers to the flow equation for $S$ (expressed in terms of $S_{\mathrm{%
int}}$) rather than to equations~(\ref{eq:SptIntLin}), (\ref{eq:GtraIntLin}) as it
would be normally. It is worth to underline, however, that the latter flow
equation for $S_{\mathrm{int}}$ is perfectly valid and may be studied for
its own sake.

Another consequence of the linear introduction of $\eta $ is the complicated
expression of the EMRO. Indeed, the linear introduction of $\eta $ is not
compatible with the translation (\ref{eq:translatAlpha1}), (\ref%
{eq:translatAlpha2}) and the solution of (\ref{eq:B}), (\ref{eq:Binit}) must
actually be considered with $\varpi _{g}=1-\eta /2$ so that, for the choice (%
\ref{eq:Gq2}), (\ref{eq:Hq2}) and with $\tilde{P}(\tilde{q}^{2})$ given by (%
\ref{eq:FormeDeP}) in which $\varpi $ is formally set equal to $0$, we get,
provided $\eta <2$:
\begin{equation}
B\left( x\right) =\left( \frac{1}{x}\right) ^{1-\eta /2}\left[ K\left(
x\right) \right] ^{2}\int_{0}^{x}u^{-\eta /2}\frac{K^{\prime }\left(
u\right) }{\left[ K\left( u\right) \right] ^{2}}\rd u\,\,,  \label{eq:BSolExpl}
\end{equation}%
which cannot be reduced to the simple form of (\ref{eq:B=P}), for arbitrary $%
K$. The result (\ref{eq:BSolExpl}) is equivalent to that obtained in
appendix D of \cite{6228} for the EMRO constructed in terms of $S_{\mathrm{%
int}}.$

\subsection{Non-linear introduction of $\protect\eta $\label%
{sec:MorrisVsWilson}}

In this section we compare two non-linear ways of introducing $\eta $: the
Morris version \cite{3357} which is based on an \textquotedblleft
anomalous\textquotedblright\ dimensional analysis and the Wilson version
extended to an arbitrary cutoff described in section \ref{sec:EWE}. We show
that the two versions are formally very close to each other but yield
different RG flow equations.

In \cite{3357}, Morris has  a priori given the field $\phi $ the dimension $%
d_{\phi }^{\left( a\right) }$ [given by (\ref{eq:dphia})] with an anomalous
part that is compensated by a cutoff function anomalously dimensioned.
Applied to our present matter, the direct consequence (the easiest to grasp) of
the Morris procedure is the modification of (\ref{eq:SptToSinptLinear}) into:%
\begin{equation}
\dot{S}=\dot{S}_{\mathrm{int}}\,,  \label{eq:SptToSinptNonLinear}
\end{equation}%
that expresses the interesting property of the flow equations for $S$ and $%
S_{\mathrm{int}}$ to have equivalent fixed points. It is important to notice
that this property is also true with the Wilson version extended to an
arbitrary cutoff function since, by construction, the renormalization of the
field exactly compensates the extra $\ell $-dependency of the cutoff
function displayed in (\ref{eq:FormeDeP}).

To get the Morris version of the flow equations for $S$ and $S_{\mathrm{int}%
} $, one must first come back to the derivation of the Polchinski-like
equation (\ref{eq:PolInt}) where the derivative w.r.t. $\ell $ is performed
at a fixed dimensioned field (RG step 1). This time, $P\left( q^{2},\ell
\Lambda _{0}\right) $ is not given by (\ref{eq:FormeDeP}) but, for
dimensional reason, by

\begin{equation}
P(q^{2},\ell \Lambda _{0})=\Lambda _{0}^{-2\varpi }\ell ^{-2\varpi }\tilde{P}%
\left( \tilde{q}^{2}\right) \,,  \label{eq:FormeDePMorris}
\end{equation}%
with $\varpi $ given by (\ref{eq:omega}) in which $n_{0}$ is
equal to zero to avoid singularities at $\left\vert q\right\vert =0$. Consequently, after the
rescaling step 2, it comes a flow equation for $S_{\mathrm{int}}$ of the
form (\ref{eq:SptIntLin}), (\ref{eq:GtraIntLin}) but with an effective $%
G\left( \tilde{q}^{2}\right) $ translated by the constant term $-\varpi $
compared to (\ref{eq:Gq2}) so that this flow equation may finally be written
under the form:%
\begin{equation}
\dot{S}_{\mathrm{int}}=\mathcal{G}_{\mathrm{tra}}\left( S_{\mathrm{int}%
}\right) +\mathcal{G}_{\mathrm{dil}}( S_{\mathrm{int}},d_{\phi
}^{\left( \mathrm{c}\right) }) -\varpi \mathcal{O}( S_{\mathrm{int}},%
\tilde{P}) \,,  \label{eq:SptIntNonLinMorris}
\end{equation}%
in which $\mathcal{O}( S_{\mathrm{int}},\tilde{P}) $ is the
expression in terms of $S_{\mathrm{int}}$ of the redundant \textquotedblleft
operator\textquotedblright\ obtained from (\ref{eq:O1}) using (\ref%
{eq:dSdPhi}), (\ref{eq:d2Sd2Phi}) (up to a field independent term):%
\begin{equation}
\mathcal{O}( S_{\mathrm{int}},\tilde{P}) =\int_{\tilde{q}}\left[
\tilde{P}\left( \tilde{q}^{2}\right) \left( \frac{\delta ^{2}S_{\mathrm{int}}%
}{\delta \tilde{\phi}_{\tilde{q}}\delta \tilde{\phi}_{-\tilde{q}}}-\frac{%
\delta S_{\mathrm{int}}}{\delta \tilde{\phi}_{\tilde{q}}}\frac{\delta S_{%
\mathrm{int}}}{\delta \tilde{\phi}_{-\tilde{q}}}\right) -\tilde{\phi}_{%
\tilde{q}}\frac{\delta S_{\mathrm{int}}}{\delta \tilde{\phi}_{\tilde{q}}}%
\right] \,.  \label{eq:OSint}
\end{equation}

The Morris-like flow equation for $S$ is then readily obtained under the
form:%
\begin{equation}
\dot{S}=\mathcal{G}_{\mathrm{tra}}\left( S\right) +\mathcal{G}_{\mathrm{dil}%
}( S,d_{\phi }^{( \mathrm{c}) }) -\mathcal{G}_{\mathrm{ren}%
}\left( S\right) \,,  \label{eq:SptMorris}
\end{equation}%
in which the three terms are defined by equations (\ref{eq:Gtra})--(\ref{eq:Gdil2}).
Notice the negative sign in front of $\mathcal{G}_{\mathrm{ren}}\left(
S\right) $ which is opposite to that of Wilson as described in section (\ref%
{sec:EWE})~--- one would observe the same change of sign in front of $\varpi $
in (\ref{eq:SptIntNonLinMorris}) compared to its Wilson-like version. This
difference is due to the fact that some aspects have formally been reversed.
On the one hand, (Wilson) one lets the coefficients of $S$ vary under the
change of scale and then renormalizes the field. On the other hand, the
\textquotedblleft renormalization step\textquotedblright\ is anticipated and
included within the cutoff function because the (useful) variation of the
coefficients of $S$ has been arbitrarily incorporated in the scaling
property of the field. Notice that in both cases the procedure amounts to
keeping  the same term of the action constant.

The two versions provide well allowed forms of ERGE, but it is worth
underlining the unusual expression of (\ref{eq:SptMorris}) that may have
consequences in practical calculations. The discussion of that issue~--- as
well as of the consequences of other considerations presented in this
article~--- is left to another publication~\cite{Berv}.

\section{Summary and conclusion}

Three different ways of introducing the anomalous dimension parameter $\eta $
in an ERGE have been considered explicitly. The non-linear procedure of the
historic first version \cite{440} (extended to an arbitrary cutoff function
in \cite{5744}) has been compared to the linear\ and non-linear versions
associated with the \textquotedblleft anomalous\textquotedblright\
dimensional analysis procedure of respectively the \textquotedblleft
modified\textquotedblright\ Polchinski version \cite{3491,6750,7205,7289}
and the Morris version \cite{3357}. Their differences in essence have been
emphasized, as well as the reasons why the non-linear versions should
provide simpler calculational frameworks. The role of the exactly marginal
redundant operator in the construction of an ERGE has also been underlined.

\section*{Acknowledgements}

It is with great pleasure that I dedicate this article to M. Kozlovskii on
his sixtieth anniversary and in remembrance of the times when we organized
some symposium in the early 1990s.

I thank H. Osborn for useful remarks.

\vspace{-5mm}

\ukrainianpart

\title%
{Рівняння точної ренормалізаційної групи  Вільсона \\ і параметр
аномальної вимірності
}%
\author{К. Бервільє}
\address{
Лабораторія математики і теоретичної фізики, UMR 7350 (CNRS),
Університет Франсуа Рабле, \\ 37200  Тур, Франція}

\makeukrtitle

\begin{abstract}
\tolerance=3000%
Параметр аномальної вимірності, введений в нелінійний спосіб у
першій історичній версії рівняння точної ренормалізаційної групи,
порівнюється з сучасною методикою. Простий вираз для точно граничного
(маргінального) надлишкового оператора слідує  з цієї нелінійності,
тоді як для того, щоб  отримати цей результат у лінійному випадку
необхідно розв'язати диференціальні рівняння першого порядку.
Висвітлено роль цього оператора в побудові рівняння потоку.
\keywords рівняння точної ренормалізаційної групи, аномальна
вимірність
\end{abstract}


\begin{thebibliography}{99}

\bibitem{440}  Wilson K.G.,  Kogut J., Phys. Rep., 1974, \textbf{12C}, 75;
\doi{10.1016/0370-1573(74)90023-4}.

\bibitem{4011}  Wegner F.J., J. Phys. C: Solid State Phys., 1974, \textbf{7}%
, 2098; \doi{10.1088/0022-3719/7/12/004}.

\bibitem{2835} Wegner F.J., %{The critical state, General aspects} ,
In: Phase Transitions and Critical
Phenomena, Vol.~{VI},   Domb C., Green M.S. (Eds.),
Academic Press, New York, 1976,  p. 7.

\bibitem{6707} Kozlovskii M.P.,  Pylyuk I.V.,  Usatenko Z.E., Phys.
Stat. Sol. B, 1996, \textbf{197}, 465; \doi{10.1002/pssb.2221970221}.

\bibitem{5175} Kozlovskii M.P.,  Patsahan O.V., Condens. Matter Phys., 2000,
\textbf{3}, 607.

\bibitem{6071} Kozlovskii M.P.,  Pylyuk I.V.,  Prytula O.O., Phys. Rev.
B, 2006, \textbf{73}, 174406; \doi{10.1103/PhysRevB.73.174406}.

\bibitem{6715}  Kozak P.R., Kozlovskii M.P.,  Usatenko Z.E., J. Phys. A:
Math. Theor., 2010, \textbf{43}, 495001; \\ \doi{10.1088/1751-8113/43/49/495001}.
%Preprint \arxiv{1005.1032}, 2010.

\bibitem{7848} Kozlovskii M.P.,  Romanik R.V., Condens. Matter Phys., 2011,
\textbf{14}, 43002; \doi{10.5488/CMP.14.43002}.
%Preprint \arxiv{1202.4569}, 2012.

\bibitem{7842}  Yukhnovskii I.R., {Phase Transitions
of the Second Order. Collective Variables Method}, World Scientific, Singapore, 1987.

\bibitem{248}  Kadanoff L.P., Physics, 1966, \textbf{2}, 263.

\bibitem{4823}  Aoki K.-I., Int. J. Mod. Phys. B, 2000, \textbf{14}, 1249;
\doi{10.1142/S0217979200000923}.

\bibitem{4595}  Bagnuls C.,  Bervillier C., Phys. Rep., 2001, \textbf{348},
91; \doi{10.1016/S0370-1573(00)00137-X}.
%\\ Preprint \arxiv{hep-th/0002034}, 2000.

\bibitem{4700}  Berges J.,  Tetradis N.,  Wetterich C., Phys. Rep., 2002,
\textbf{363}, 223; \doi{10.1016/S0370-1573(01)00098-9}.
%Preprint \arxiv{hep-ph/0005122}, 2000.

\bibitem{5141}  Polonyi J., Cent. Eur. J. Phys., 2003, \textbf{1}, 1;
\doi{10.2478/BF02475552}.
%Preprint \arxiv{hep-th/0110026}, 2011.

\bibitem{5604}  Delamotte B.,  Mouhanna D.,  Tissier M., Phys. Rev. B, 2004,
\textbf{69}, 134413; \doi{10.1103/PhysRevB.69.134413}.
%Preprint \arxiv{cond-mat/0309101}, 2003.

\bibitem{6022}  Pawlowski J.M., Ann. Phys., 2007, \textbf{322}, 2831;
\doi{10.1016/j.aop.2007.01.007}.
%Preprint \arxiv{hep-th/0512261}, 2005.

\bibitem{6156}  Delamotte B., %{An Introduction to the Nonperturbative Renormalization Group},
In: {Order, Disorder and Criticality. Advanced Problems
of Phase Transition Theory},  Vol.~2, Holovatch~Yu. (Ed.), World Scientific, Singapore, 2007,  p.~1.
%[arxiv:cond-mat/0702365].

\bibitem{6723}  Kopietz P.,  Bartosch L.,  Sch\"{u}tz F., Lecture Notes in
Physics, 2010, \textbf{798}; \doi{10.1007/978-3-642-05094-7}.

\bibitem{6699}  Rosten O.J., Phys. Rep., 2012, \textbf{511}, 177;
\doi{10.1016/j.physrep.2011.12.003}.
%Preprint \arxiv{1003.1366}, 2010.

\bibitem{4245}  Seide S.,  Wetterich C., Nucl. Phys. B, 1999, \textbf{562},
524; \doi{10.1016/S0550-3213(99)00545-3}.
%\\ Preprint \arxiv{cond-mat/9806372}, 1998.

\bibitem{4716}  Tissier M.,  Delamotte B.,  Mouhanna D., Phys. Rev. Lett.,
2000, \textbf{84}, 5208; \doi{10.1103/PhysRevLett.84.5208}.
%Preprint \arxiv{cond-mat/0001350}, 2000.

\bibitem{6709}  Machado T.,  Dupuis N., Phys. Rev. E, 2010, \textbf{82},
041128; \doi{10.1103/PhysRevE.82.041128}.
%Preprint \arxiv{1004.3651}, 2010.

\bibitem{4753}  Latorre J.I.,  Morris T.R., J. High Energy Phys., 2000, \textbf{11}, 004;
\doi{10.1088/1126-6708/2000/11/004}.
% Preprint \arxiv{hep-th/0008123}.

\bibitem{3358} Morris T.R., Phys. Lett. B, 1994, \textbf{334}, 355;
\doi{10.1016/0370-2693(94)90767-6}.
%Preprint \arxiv{hep-th/9405190}, 1994.

\bibitem{2520} Morris T.R., Int. J. Mod. Phys. A, 1994, \textbf{9}, 2411;
\doi{10.1142/S0217751X94000972}.
%Preprint \arxiv{hep-ph/9308265}, 1993.

\bibitem{3550} Morris T.R., Nucl. Phys. B, 1996, \textbf{458}, 477;
\doi{10.1016/0550-3213(95)00541-2}.
%Preprint \arxiv{hep-th/9508017}, 1995.

\bibitem{4619}  Sumi J.-I.,  Souma W.,  Aoki K.-I.,  Terao H.,  Morikawa K.,
%{Scheme dependence of the Wilsonian effective action and sharp cutoff limit of the flow equation},
Preprint \arxiv{hep-th/0002231}, 2000.

\bibitem{3357} Morris T.R., Phys. Lett. B, 1994, \textbf{329}, 241;
\doi{10.1016/0370-2693(94)90767-6}.
%Preprint \arxiv{hep-ph/9403340}, 1994.

\bibitem{3993} Morris T.R., Prog. Theor. Phys. Suppl., 1998, \textbf{131},
395; \doi{10.1143/PTPS.131.395}.
%Preprint \arxiv{hep-th/9802039}, 1998.

\bibitem{4436}  Ellwanger U., Z. Phys. C, 1994, \textbf{62}, 503;
\doi{10.1007/BF01555911}.
%Preprint \arxiv{hep-ph/9308260}, 1993.

\bibitem{5744}  Bervillier C., Phys. Lett. A, 2004, \textbf{332}, 93;
\doi{10.1016/j.physleta.2004.09.037}.
%Preprint \arxiv{hep-th/0405025}, 2004.

\bibitem{6750}  Rosten O.J., J. Phys. A: Math. Theor., 2011, \textbf{44},
195401; \doi{10.1088/1751-8113/44/19/195401}.
%\\ Preprint \arxiv{1010.1530}, 2010.

\bibitem{7289}  Osborn H.,  Twigg D. E., Ann. Phys., 2012, \textbf{327},
29; \doi{10.1016/j.aop.2011.10.011}.
%Preprint \arxiv{1108.5340}, 2011.

\bibitem{7205}  {Rosten O.J.,
%{Relationships Between Exact RGs and some Comments on Asymptotic Safety},
Preprint \arxiv{1106.2544}, 2011.}

\bibitem{354}  Polchinski J., Nucl. Phys. B, 1984,  \textbf{231}, 269;
\doi{10.1016/0550-3213(84)90287-6}.

\bibitem{3491}  Ball R.D.,  Haagensen P.E.,  Latorre J.I.,  Moreno E.,
Phys. Lett. B, 1995, \textbf{347}, 80; \doi{10.1016/0370-2693(95)00025-G}.
%Preprint \arxiv{hep-th/9411122}, 1994.

\bibitem{6228}  O'Dwyer J.P.,  Osborn H., Ann. Phys., 2008, \textbf{323}, 1859; \doi{10.1016/j.aop.2007.10.005}.
%Preprint \arxiv{0708.2697v2}, 2007.

\bibitem{4405}  {Riedel E.K.,  Golner G.R.,  Newman K.E., Ann. Phys., 1985,
\textbf{161}, 178; \doi{10.1016/0003-4916(85)90341-0}.}

\bibitem{Berv} Bervillier C. (unpublished).

\bibitem{3912} Golner G.R.,
%Exact renormalization group flow equations for free energies and N-point functions in uniform
%external fields},
Preprint \arxiv{hep-th/9801124}, 1998.

\bibitem{4421}  Bell T.L.,  Wilson K.G., Phys. Rev. B, 1975, \textbf{11},
3431; \doi{10.1103/PhysRevB.11.3431}.
\end{thebibliography}
\end{document}